\documentclass[structabst]{aa}
\usepackage{natbib}
\bibpunct{(}{)}{;}{a}{}{,}

\usepackage{graphicx}
\usepackage{multirow}
\usepackage{subfigure}
\usepackage{lscape}

\newcommand{\tth}{2$\theta$ }
\newcommand{\fig}[1]{Figure~\ref{#1}}
\newcommand{\dgr}{$^\circ$ }
\newcommand{\ca}{CaSiO$_3$ }

\author{Sarah J. Day\inst{\ref{inst1}, \ref{inst2}}
 \and Stephen P. Thompson\inst{\ref{inst2}} \and Julia E. Parker\inst{\ref{inst2}} \and Aneurin Evans\inst{\ref{inst1}}}

\institute{Astrophysics Group, Keele University, Keele, Staffordshire, UK, ST5 5BG \label{inst1}
\and Diamond Light Source, Harwell Science and Innovation Campus, Chilton, Didcot, Oxon OX11 0QX \label{inst2}}

\title{Non-aqueous formation of the calcium carbonate polymorph vaterite: astrophysical implications}

\abstract 
{}
{To study the formation of calcium carbonate, through the solid-gas interaction of amorphous Ca-silicate with
gaseous CO2, at elevated pressures, and link this to the possible presence of calcium carbonate in a number of circumstellar and
planetary environments.}
{We use in-situ synchrotron X-Ray powder diffraction to obtain detailed structural data pertaining to the formation of the crystalline calcium carbonate phase vaterite and its evolution with temperature.} 
{We found that the metastable calcium carbonate phase vaterite was formed alongside calcite, at elevated CO$_2$ pressure, at room temperature and subsequently remained stable over a large range of temperature and pressure.} 
{We report the formation of the calcium carbonate mineral vaterite whilst attempting to simulate carbonate dust grain formation in astrophysical environments. This suggests that vaterite could be a mineral component of carbonate dust and also presents a possible method of formation for vaterite and its polymorphs on planetary surfaces.}

\begin{document}
\titlerunning{Formation of vaterite: astrophysical implications}
\maketitle

\section{Introduction}
\subsection{Carbonates in astrophysical environments}

On Earth, carbonates are the main constituent in chemically precipitated sedimentary rocks (e.g. limestone, dolomite) and are produced organically by marine organisms whose shells or skeletal structures are built out of calcium carbonate; however, the inorganic formation of carbonates in terrestrial environments has been found to occur exclusively through the interaction of silicate minerals with liquid water containing dissolved CO${_3^{2-}}$. However, features indicative of carbonate minerals have been detected in a wide variety of astrophysical environments, from planetary bodies within the Solar System \citep{ehlmann08} to dust particles surrounding stellar objects \citep{kemper02,toppani05, chiavassa05}.

\subsubsection{Circumstellar Environments}

Carbonates have been detected in a number of circumstellar environments, exhibiting far-infrared (FIR) features around 62$\mu$m and 92$\mu$m. The 92$\mu$m feature has been identified as calcite (CaCO$_3$), being the only dust species known to have a strong feature at this wavelength. Dolomite (CaMgCO$_3$) has a band at $\sim$62$\mu$m, and has been suggested as a possible dust species; however due to other strong features of H$_2$O and diopside (CaMgSi$_2$O$_6$) between 62--65$\mu$m this identification is tenuous. 
Mg-rich and Fe-rich carbonates such as magnesite and siderite do not exhibit strong features in the far infrared and therefore cannot be confirmed as dust components.

\cite{kemper02} report the detection of carbonate features in the planetary nebulae (PNe) NGC6302 and NGC6537. They argue that the carbonates in these environments require formation by non-aqueous routes, such as gas-phase condensation or processes on grain surfaces. \cite{ceccarelli02} have also identified carbonate features in the proto-star NGC1333--IRAS4, and they too argue for a non-aqueous origin. \cite{chiavassa05} have identified carbonate features in a range of pre-main sequence environments, from ``Class 0'' objects to Herbig AeBe stars, and identify a need for experimental evidence of non-aqueous formation mechanisms. These observations demonstrate that carbonates are present in both pre-main sequence and highly evolved environments.

A number of non-aqueous formation routes have been proposed to explain the presence of carbonates in circumstellar environments. These include: the processing of CO$_2$--rich ices through catalytic reactions on hydrated silicate surfaces \citep{ceccarelli02}, direct condensation at high temperatures from gas rich in CaO and CO$_2$ \citep{toppani05}, and solid-gas interaction of silicate grains and hot, gaseous CO$_2$ \citep{kemper02}. However, \cite{ferrarotti05} calculated that direct condensation of carbonate grains would not be viable in such environments due to the low temperature (\textless 800K) required for carbonates to precipitate from the gas phase. At distances from the central star implied by these temperatures the outflowing gas would be significantly diluted, suppressing the formation of less stable dust species such as carbonates. In pre-main sequence environments carbonates could form through processing on larger planetesimals within the proto-planetary disk. Collisions of such bodies would provide a source for the carbonate dust observed in these systems. \cite{rietmeijer08} support the carbonate formation method suggested by \cite{kemper02}. They have shown experimentally that solid, amorphous CaSiO$_3$, formed though vapour phase condensation, as a deep metastable eutectic composition which could easily crystallise as a refractory mineral in astrophysical environments. \cite{rietmeijer08} proposed that these amorphous, Ca-rich silicates could then react with gaseous CO$_2$ to produce calcium carbonate (CaCO$_3$) via the simple, single-step reaction:
\begin{equation}
\mbox{CaSiO}_3 + \mbox{CO}_2 \rightarrow \mbox{CaCO}_3 + \mbox{SiO}_2  .
\label{carbonation}
\end{equation}

Additionally, \cite{chiavassa05} emphasise the fact that both PNe and young stellar objects are known sources of X-Ray emission and suggest that carbonates could form in both environments through the interaction of X-Rays with water-ice on silicate dust grains, increasing the mobility of H$_2$O close to that of liquid water.

\subsubsection{Planetary Environments}
Carbonates have long been known to be present within chondritic meteorites \citep{rubin97} and have typically been used to infer the presence of liquid water. In the case of meteorites, the presence of carbonate phases is believed to be the result of aqueous alteration of silicate minerals on a larger parent body \citep{fredriksson88}.
A variety of carbonates are known to be present on larger bodies in the solar system, primarily Mars and Venus \citep{ehlmann08, viviano12}; and in these cases their presence is also often associated with aqueous alteration during the planets' history.
 Mars and Venus are both known to have abundant CO$_2$ within their atmospheres, albeit at vastly different atmospheric pressures, and it has been suggested that chemical weathering could lead to the formation of carbonates through reaction (\ref{carbonation}) \citep{visscher09}. This solid-gas reaction of Ca-rich silicates with CO$_2$ has been suggested as a possible method for regulating the atmospheric CO$_2$ on Venus \citep{visscher09}, with calcium silicates on its surface reacting with the high surface CO$_2$ surface pressure of 90 bar, at a temperature of 750~K \citep{wood68}; however, this is currently a matter of debate \citep{treiman12} and assumes that the abundance of Ca-rich silicates exposed on the surface of Venus is sufficiently high. It has also been argued that this would not be a viable buffer of CO$_2$ and there is doubt about the abundance of wollastonite (CaSiO$_3$) on the surface of Venus to allow this reaction to occur \citep{treiman12}. 

Carbonate features have also been reported on the surface of Mars using the Compact Reconnaissance Imaging Spectrometer for Mars (CRISM) on the Mars Reconnaissance Orbiter \citep{ehlmann08, viviano12}. These carbonates have been found in localised areas on Mars, with little evidence for carbonate minerals elsewhere on the surface of the planet. \cite{vanberk12} suggests that in Mars' history the partial pressure of CO$_2$ was much higher than it is today, and that the carbonates could have been formed through exposure to CO$_2$ at a partial pressure of 1 bar over a long period of time (\textgreater 10$^4 - 10^5$ years).

The Martian meteorite ALH884001 was found to contain spherical carbonate structures that, due to their well defined shape, were initially proposed as evidence for past life on Mars \citep{mckay96}. However, an alternative interpretation proposed by \cite{vecht00} was that the spherical features were indicative of vaterite (metastable CaCO$_3$, see Section~\ref{vaterite}) known to grow in a distinctive spherical habitat, and although no longer of vateritic composition, they were likely formed inorganically by more stable carbonate phases pseudo-morphing an earlier vaterite assemblage.

\subsection{Vaterite}{\label{vaterite}

The three predominant anhydrous calcium carbonate mineral phases are calcite, aragonite and vaterite. Of these, calcite is the most stable, with aragonite and vaterite being metastable phases, existing only under a narrow range of temperature and pressure conditions. Calcite is therefore the most abundant and as such, it is generally calcite that is attributed to carbonate features found within astrophysical environments \citep[e.g.][]{ceccarelli02,chiavassa05}. However, vaterite has been found in a selection of enstatite achondrite meteorites \citep{dufresne62, okada81, vecht00}.

Naturally forming vaterite is rarely found on Earth, predominantly occurring in association with organic tissue \citep{sutor69} and bacteriological biomineralisation products \citep[e.g.][]{rodriguez03, sanchez03, falini05, cacchio04, demuynck08, chen09, zamarreno09} and freshwater mollusc shells and pearls \citep[e.g.][]{wehrmeister07, jacob08, soldati08}. Naturally occuring vaterite has been discovered in small amounts within zones of contact metamorphism \citep{mcconnell60, kolodny74}, drilling mud \citep{friedman93} and in stagnant, natural waters, forming as a result of spontaneous precipitation \citep[e.g.][]{rowlands71, grasby03}. 

Vaterite can be inorganically synthesized in the laboratory through a number of processes, including precipitation of a calcium carbonate gel from concentrated calcium and carbonate solutions \citep{andreassen05}, through the decomposition of the hydrated carbonate phase ikaite \citep[CaCO$_3$.6H$_2$O;][]{tang09} or stabilised via biomimetic-based precipitation in the presence of organic additives \citep[e.g.][]{thompson11a,thompson11b}, although these processes are often involved and, though reproducible, are often sensitive to variations in the preparation conditions. Once formed, vaterite is metastable and very soluble, more so than the calcite and aragonite polymorphs, converting into calcite in less than 25 hours at room temperature \citep{silk70} when in contact with water. In the absence of water it has been found to transform into calcite at temperatures between 730~K and 840~K \citep{subba73,peric96}.

The crystal structure of vaterite has been studied in detail over the last few decades; however the rarity of vaterite in nature and its reluctance to form large single crystals has lead to a number of uncertainties regarding its crystallographic structure (see Table~\ref{cryststruct} for a summary). The main reports of the vaterite crystal structure come from \cite{kamhi63} and \cite{meyer60, meyer69}. \cite{meyer60,meyer69} suggests both a hexagonal and orthorhombic unit cell, whereas \cite{kamhi63} reports a hexagonal unit cell with a prominent hexagonal pseudo-cell. These cells can be linked, and the orthorhombic cell of Meyer can be related to the hexagonal Kahmi cell via a transformation of lattice parameters \citep[see discussion by][]{tang09}. More recently, \cite{lebail11}, in an attempt to fit single-crystal data of vaterite, propose an $Ama$ 2 space group using a model based on crystal microtwinning (3 orthorhombic domains rotated by 120\dgr to produce pseudohexagonal symmetry). Additionally, \cite{mugnaioli12} have used electron diffraction techniques, allowing the collection of diffraction data from single nanocrystals, to propose a monoclinic structure, with a $C$2/$c$ space group. However, carbonate anions in vaterite
exhibit disorder such that their locations are still a matter of debate \citep{medeiros07, gebauer09, jacob09, wehrmeister10}. Density Functional Theory (DFT) calculation by \cite{demichelis12} suggest the previous {\it Pbnm} and $P$6$_{5}$22 structures could in fact represent unstable transition states leading towards a more stable structure with $P$3$_{2}$21 symmetry. The DFT calculations suggest there are at least three distinct minima whose energies and activation barriers for interconversion are all within the accessible range of thermal energy at room temperature. \cite{wang09b}, however, suggest the $P$6$_5$22 space group for a stable, fully ordered vaterite structure, obtained using first principles calculations and molecular dynamic simulations. Therefore, the uncertainty surrounding the structure of vaterite observed in different experiments could be due to differing combinations of structures involving different symmetries.

\begin{table}

\caption{Comparison of proposed crystal structures for vaterite from the literature. \label{cryststruct}}

\begin{tabular}{l l l l l}
\hline
\multicolumn{1}{c}{Reference} & \multicolumn{3}{c}{Unit Cell Parameters (\AA)} & \multicolumn{1}{p{1.55cm}}{Space Grp.} \\ \hline
 & &  \\
Demichelis (2012) & $a$ = 7.12& & $c$ = 25.32 &$P$3$_2$21 \\
Mugnaioli (2012) & $a$ =12.17& $b$ = 7.12 & $c$ = 9.47 & $C$2/$c$ \\
Wang (2009) & $a$ = 7.29 & $b$ = 7.29 & $c$ = 25.30 & $P$6$_5$22 \\ 
Le Bail (2009)& $a$ = 8.47 & $b$ = 7.16 & $c$ = 4.12 & $Ama$2 \\
Kahmi (1963) & $a$ = 7.16 & & $c$ = 16.98 & $P$6$_3$/$mmc$\\
Kamhi (1963) & $a'$= 4.13 & & $c'$ = 8.49 & $P$6$_3$/$mmc$ \\
Meyer (1960)&  $a$ = 4.13 & $b$ = 7.15 & $c$ = 8.48 &  $Pbnm$\\ \hline
\end{tabular}

\end{table}

In this paper we report on the formation of vaterite, through a solid-gas reaction of amorphous calcium silicate powders (CaSiO$_3$) and gaseous CO$_2$ at elevated pressure (6--40 bar) and temperature (273 -- 1223~K), with the intention of demonstrating how carbonate phases might form in non-aqueous astrophysical environments. We focus on carbonation of pure Ca-silicates especially, as Mg-rich carbonates do not fit the observed 92$\mu$m feature, indicative of carbonates, in astronomical spectra \citep{kemper02}.

\section{Experimental \label{expt}}

An amorphous calcium-rich silicate was produced via a sol gel method \citep[described in detail by][]{thompson12a}, in which the metal salts CaCl$_2$ and Na$_2$SiO$_3$ were combined in 0.1M solutions to form samples having the stoichiometric composition \ca. In order to prevent carbonation of the final samples through the interaction with atmospheric CO$_2$, they were stored in gel form under demineralised water until required, at which point the gels were dried in a Carbolite HVT vacuum furnace (P \textless 10$^{-4}$ mbar), initially purged with Nitrogen. The gels were dried at a temperature of 323~K for approximately 1 hour, producing very fine-grained powders (e.g. Figure ~\ref{SEMvac}) with an average grain size of \textless 10$\mu$m. Upon removal from the furnace samples were immediately stored in sealed glass vials under Argon, prior to being loaded into the experimental cell. 

\begin{figure}
\begin{center}
 \includegraphics[scale=0.4]{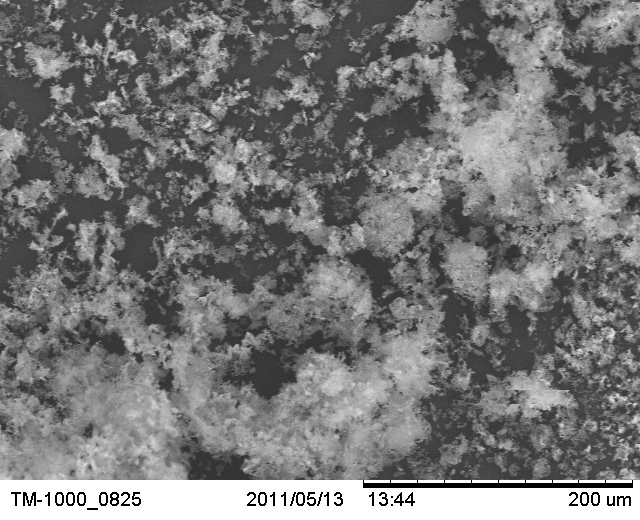}
\end{center}
\caption{Scanning electron microscope image of the final powder product with composition CaSiO$_3$, vacuum dried at 323 K. \label{SEMvac}}
\end{figure}

In-situ synchrotron X-ray powder diffraction (SXPD) measurements were taken on Beamline I11 \citep{thompson09} at the Diamond Light Source, UK using a position sensitive detector (PSD) allowing for fast, real-time data collection \citep{thompson11a}. The I11 beamline receives low divergence, high intensity X-rays from a 22-pole undulator source, monochromatic X-rays are selected by a double bounce Si(111) monochromator, coupled with double bounce harmonic rejection mirrors. The PSD comprises 18 pixellated Si strip detection modules tiled to give a 90\dgr measurement arc. The X-ray wavelength, calibrated against a NIST SRM 640c standard Si powder, was 0.825590\AA.

A high pressure gas cell \citep[][see Figure~\ref{gascell}]{thompson12b} was used in order to expose the sample to high purity (99.98\%) CO$_2$ at varying pressures (1--40 bar), and heating of the sample was performed using a Cyberstar hot air blower (Ramp rate: 10$^\circ$/min up to 1273~K) located beneath the sample capillary. Powder samples were loaded into 0.7mm diameter quartz capillaries (maximum pressure 50 bar) or 0.79mm sapphire tubes (maximum pressure 100 bar) and held in place with a quartz wool plug. It should be noted that, while the sapphire tubes are straight and have a constant diameter along their length, the quartz capillaries are tapered towards to the (closed) end of the capillary; consequently they are usually smaller than their nominal diameter. It was therefore found necessary to apply greater pressure to the powder to fill the quartz capillaries, resulting in tighter packing; as discussed below this has implications for the gas pressure required to initiate the reaction. The samples were mounted into the high pressure cell, aligned on the diffractometer and connected to the gas control system \citep{parker11}. Prior to connecting the sample to the system the gas supply line was pumped down to vacuum to remove air from the system and CO$_2$ was then injected to 1 bar. 

Once the sample was mounted the CO$_2$ pressure was increased to 6 bar, to ensure the adequate gas penetration through the length of the sample to position of the X-ray beam. However, due to the slightly smaller diameter of the quartz capillary and the tighter packing of the sample within the capillary, it was determined that an initial pressure of 20 bar was required in order to obtain adequate gas penetration through the sample and observe a reaction at the beam position.
 
Due to the tighter packing of the sample in the quartz capillaries, the gas is prevented from penetrating through the complete length of the capillary, instead reacting only with the material at the entrance to the capillary, outside of the section covered by the beam. In this case, the additional pressure required to observe a reaction when using the quartz capillary is not believed to be due to any difference in the sample but merely represents the additional pressure required to allow the gas to diffuse through the full length of the capillary. \cite{thompson12a} reported the carbonation of the same amorphous samples in a vacuum desiccator using ammonium carbonate as the source of CO$_2$, supporting the fact that the need for higher pressure in this study is solely due to the experimental setup, and is not necessary for such a reaction to occur.

The sample was heated steadily from room temperature (RT; 298~K) to 1223~K and data were collected at 5~K intervals. To provide sufficient powder averaging during data collection, the sample capillary was rocked $\pm$15$^{\circ}$ about its length. Every scan with the PSD took 10 seconds. For subsequent samples the pressure was increased at ambient temperature, in steps of 10 bar, up to a maximum pressure of 40 bar.

\begin{figure}
\begin{center}
 \includegraphics[scale=0.5]{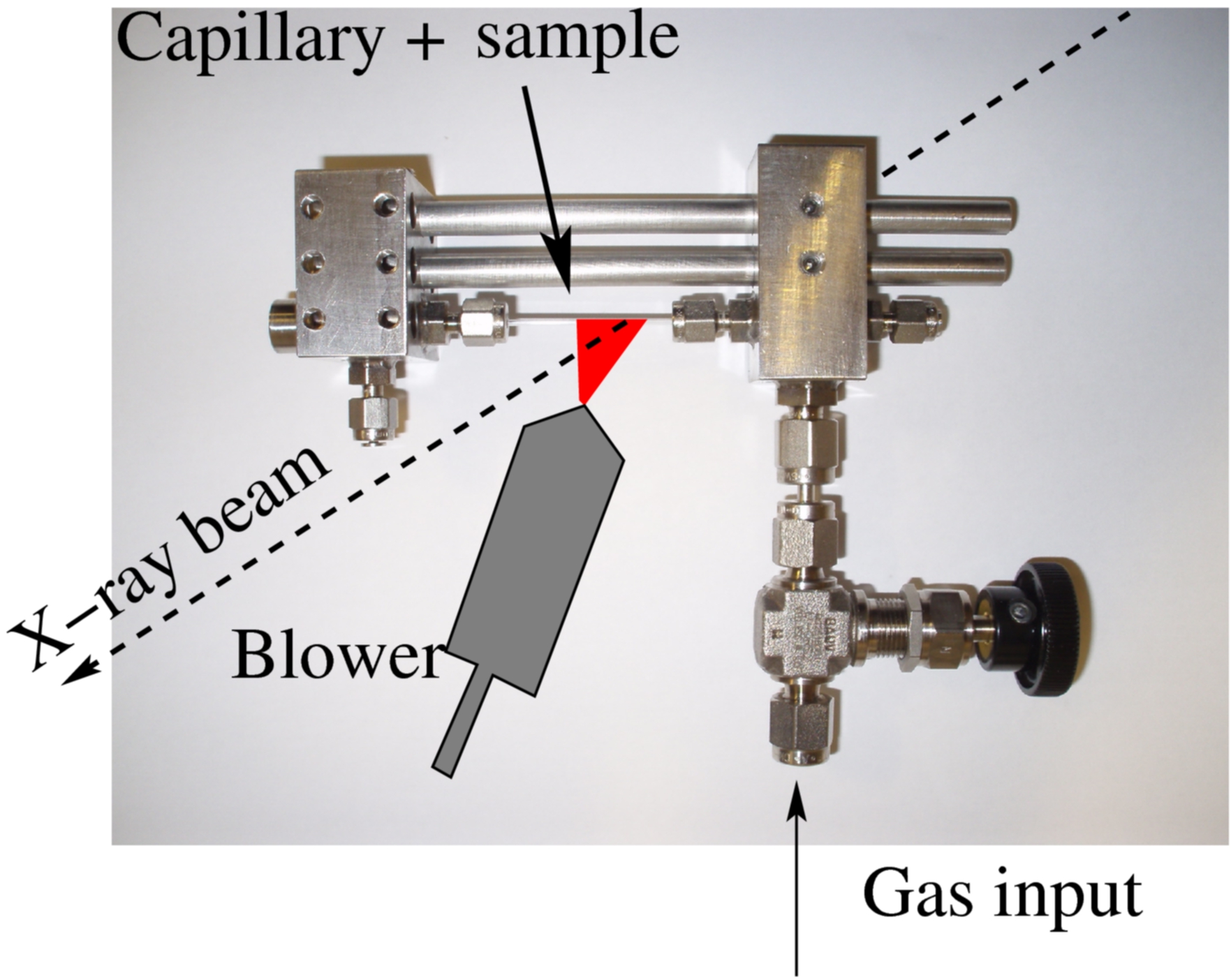}
\end{center}
\caption{High pressure gas cell with sapphire tube. X-ray beam is at a right angle to the capillary. \label{gascell}}
\end{figure}

\section{Results}

\begin{figure*}
\begin{center}
\includegraphics[scale=0.5]{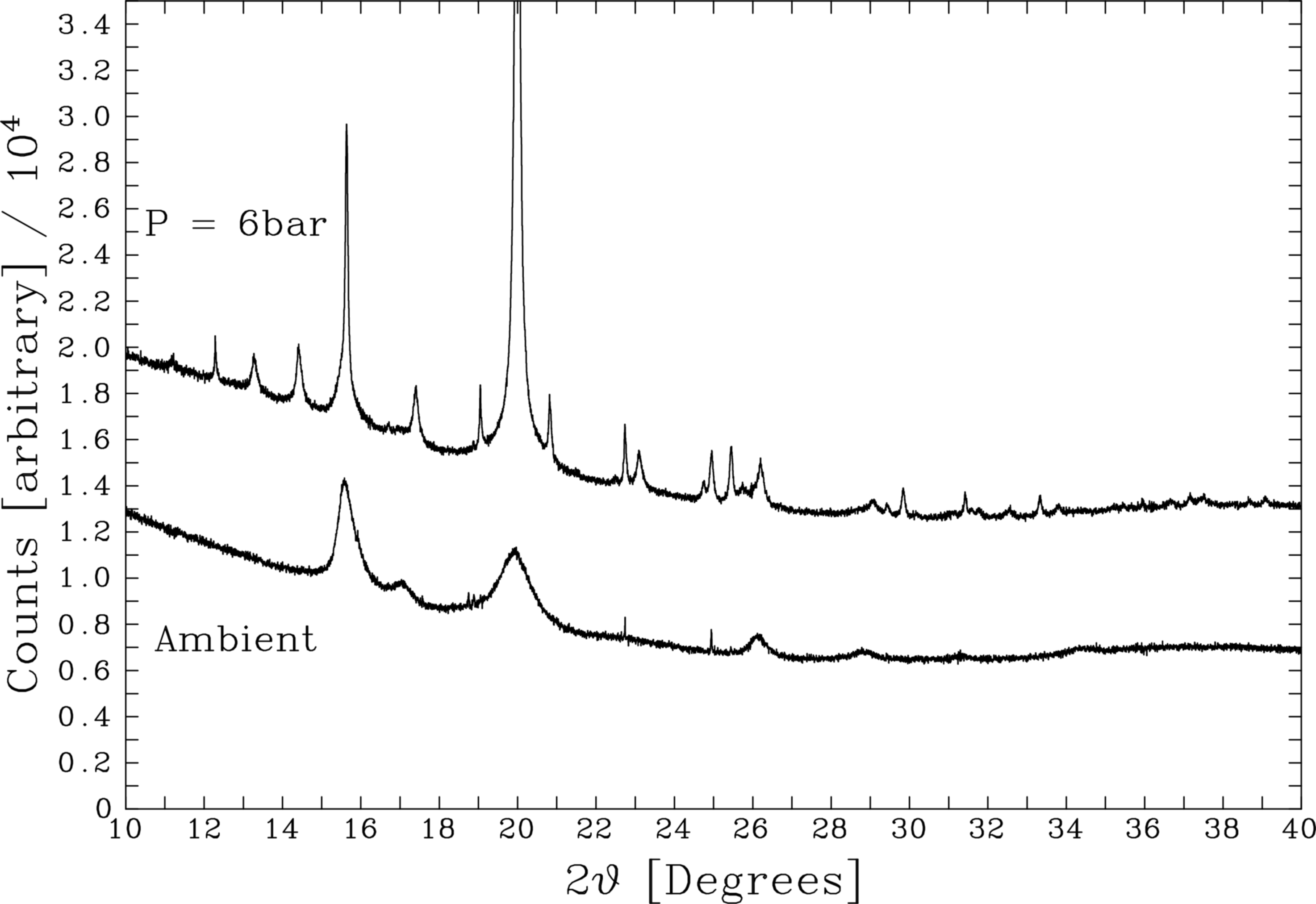}
\end{center}
\caption{A comparison of the diffraction patterns for the powdered sample of CaSiO$_3$ before (lower curve) and after (upper curve) the injection of CO$_2$. The strong feature at 19.97\dgr 2$\theta$ is due to the sapphire tube. Patterns are offset in the y-axis. \label{vateriteinitial}}
\end{figure*}

Upon initial injection of CO$_2$, at a pressure of 6 bar, into the larger diameter sapphire tube, the formation of a crystalline carbonate phase was observed almost instantaneously. Figure ~\ref{vateriteinitial} shows a comparison of the initial SXPD pattern of the CaSiO$_3$ sample under vacuum at ambient temperature and the first SXPD scan taken as the CO$_2$ was opened to the sample. Injection of CO$_2$ triggered the formation of a crystalline phase.

\begin{figure*}
\begin{center}
\includegraphics[scale=0.46]{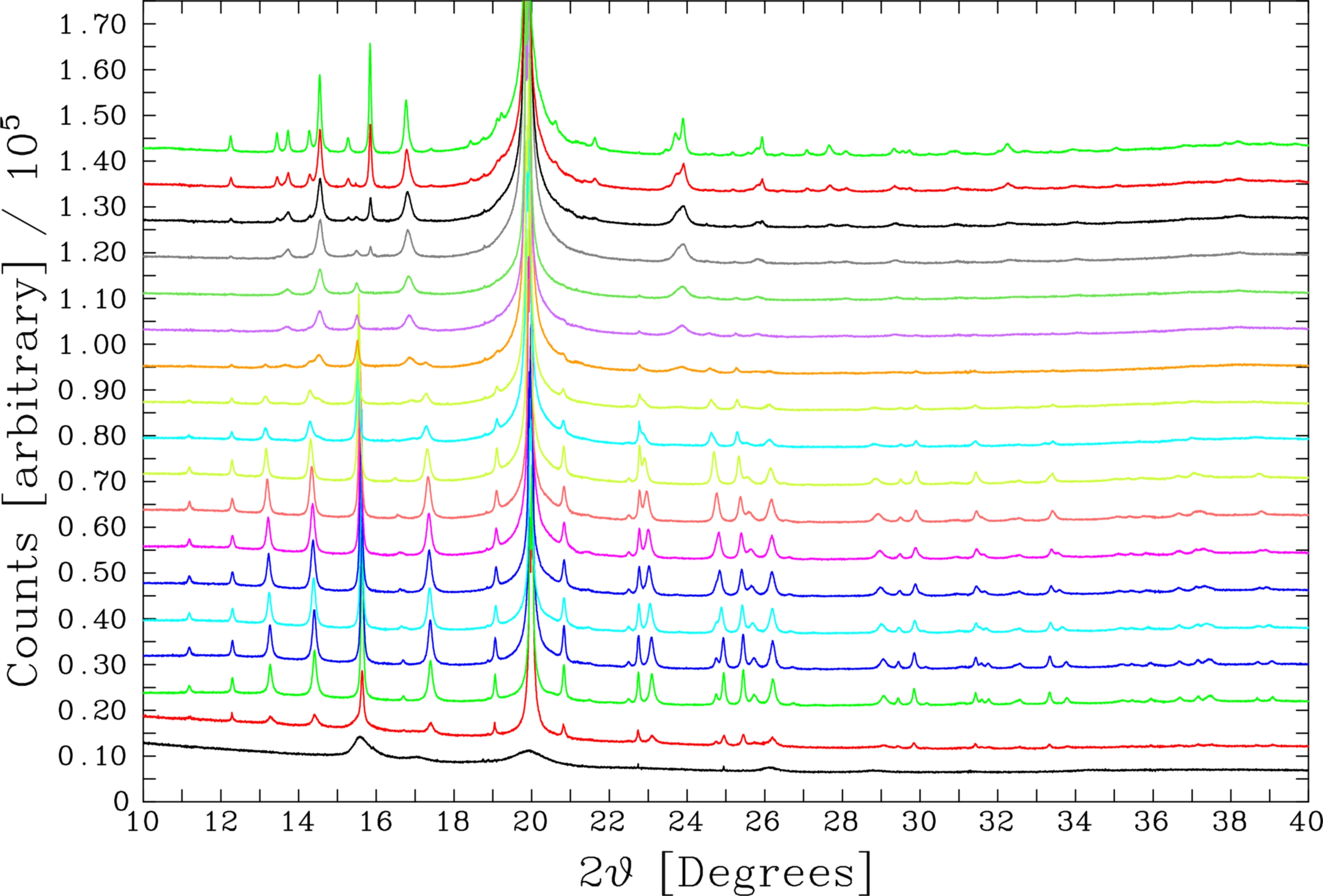}
\end{center}
\caption{Thermal evolution of CaSiO$_3$ exposed to CO$_2$ at 6 bar whilst being heated from room temperature (bottom scan) to 1173 K (top scan). Patterns have been offset on the y-axis for clarity and temperature is increasing in steps of 50~K from bottom to top. The strong feature at 19.97 \tth is due to the sapphire tube. \label{sapphthermalevo}}
\end{figure*} 

Once crystallised the sample exhibits multiple strong, well-resolved features, suggesting a well ordered crystalline structure. The distinctive groups of features at 10\dgr-- 14.5\dgr \tth, 22\dgr-- 24\dgr \tth and 25\dgr-- 27\dgr \tth are attributed to vaterite, while the remaining features at 12.2$^{\circ}$, 15.8$^{\circ}$, 19.1$^{\circ}$ and 20.9\dgr \tth are due to calcite. Peak search-match software Match!\footnote{Crystal Impact http://www.crystalimpact.com/match/} suggested that the pattern could be fit by a mixture of the calcium carbonate phases vaterite and calcite. 

The pressure was held at 6 bar at RT for approximately 30 minutes. During this time the structure continued to evolve, with the peaks steadily growing in intensity; this was accompanied by a slow but steady decrease in the CO$_2$ pressure at the sample. We should note however, that we cannot rule out the presence of an amorphous silicate component remaining in the sample; evidence suggests that the amorphous silicate component anneals at higher temperatures, or after prolonged exposure ($\sim$5 hours) at T \textgreater 1220~K \citep{fabian00}.

 \begin{figure*}[ht]
\begin{center}
\includegraphics[scale=0.45]{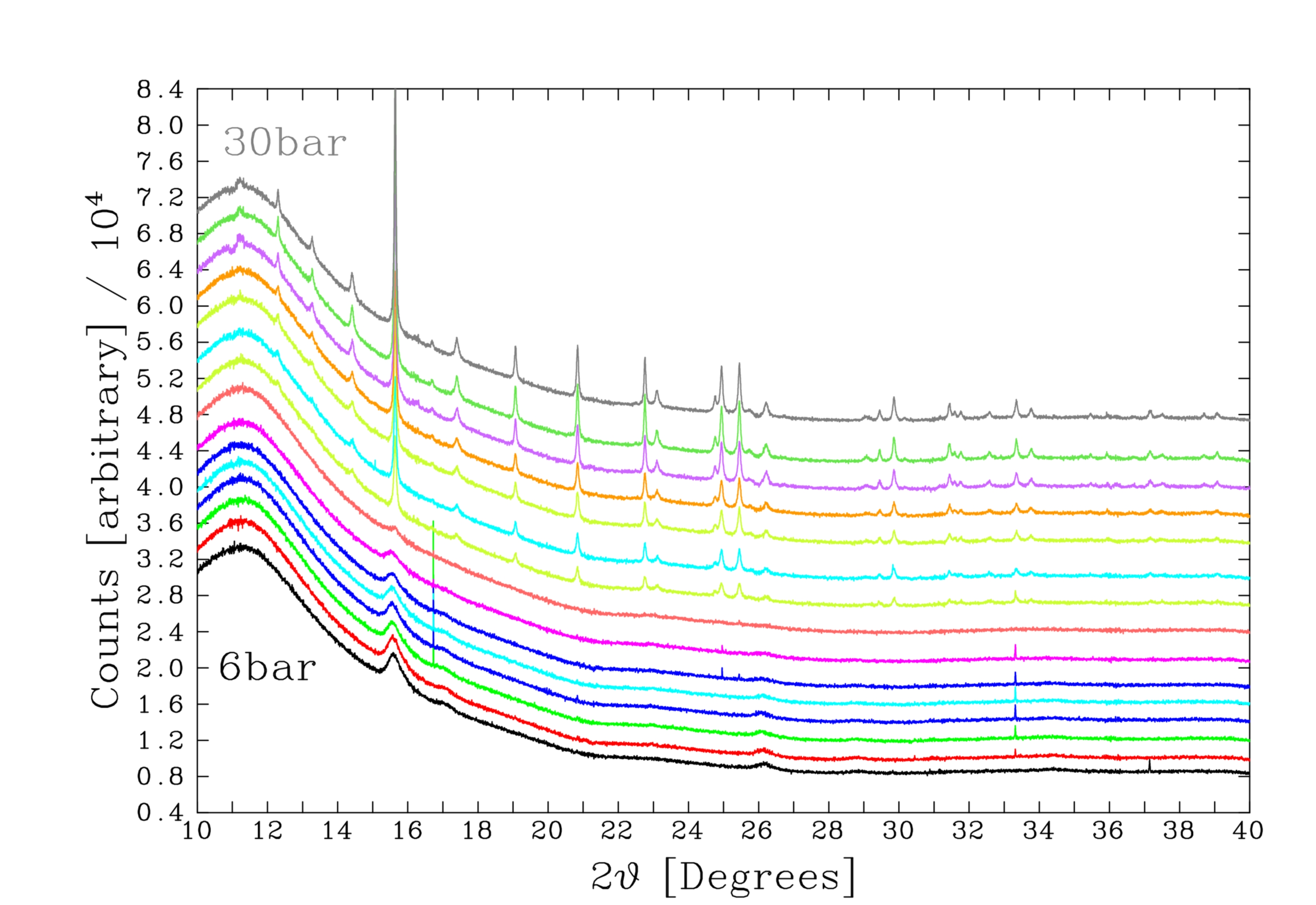}
\end{center}
\caption{Evolution of CaSiO$_3$ in quartz capillary with increasing pressure of CO$_2$. Patterns have been offset on the y-axis for clarity and high backround is due to the quartz capillary. Pressure increases upwards from 6 bar to a maximum of 30 bar. \label{quartzpressevo}}
\end{figure*} 

\subsection{Thermal Evolution}

Heating of the sample was initiated after 30 minutes at a pressure of 6 bar. During this time the CO$_2$ pressure dropped steadily, an indication of the reaction still occurring, and was topped up to 6 bar when the pressure dropped below 4 bar (after approximately 2.5 hours). The experiment was then left to run overnight, with scans being taken at 5~K intervals (approximately every 3 mins) between RT and 1223~K, during which the pressure had dropped and settled at 2.6 bar. The results of this are plotted in \fig{sapphthermalevo} presenting a selection of patterns at roughly 50~K increments. These show that, following the initial formation of the carbonate phases, they remain relatively stable during heating until the temperature exceeds 753~K, at which point the carbonate phases start to break down and the sample begins to anneal, finally crystallising, at T \textgreater 1083~K, to wollastonite (CaSiO$_3$). This is discussed in greater detail in \textsection ~\ref{sapph}.

\begin{center}
\begin{figure*}

\subfigure[]{\includegraphics[scale=0.55]{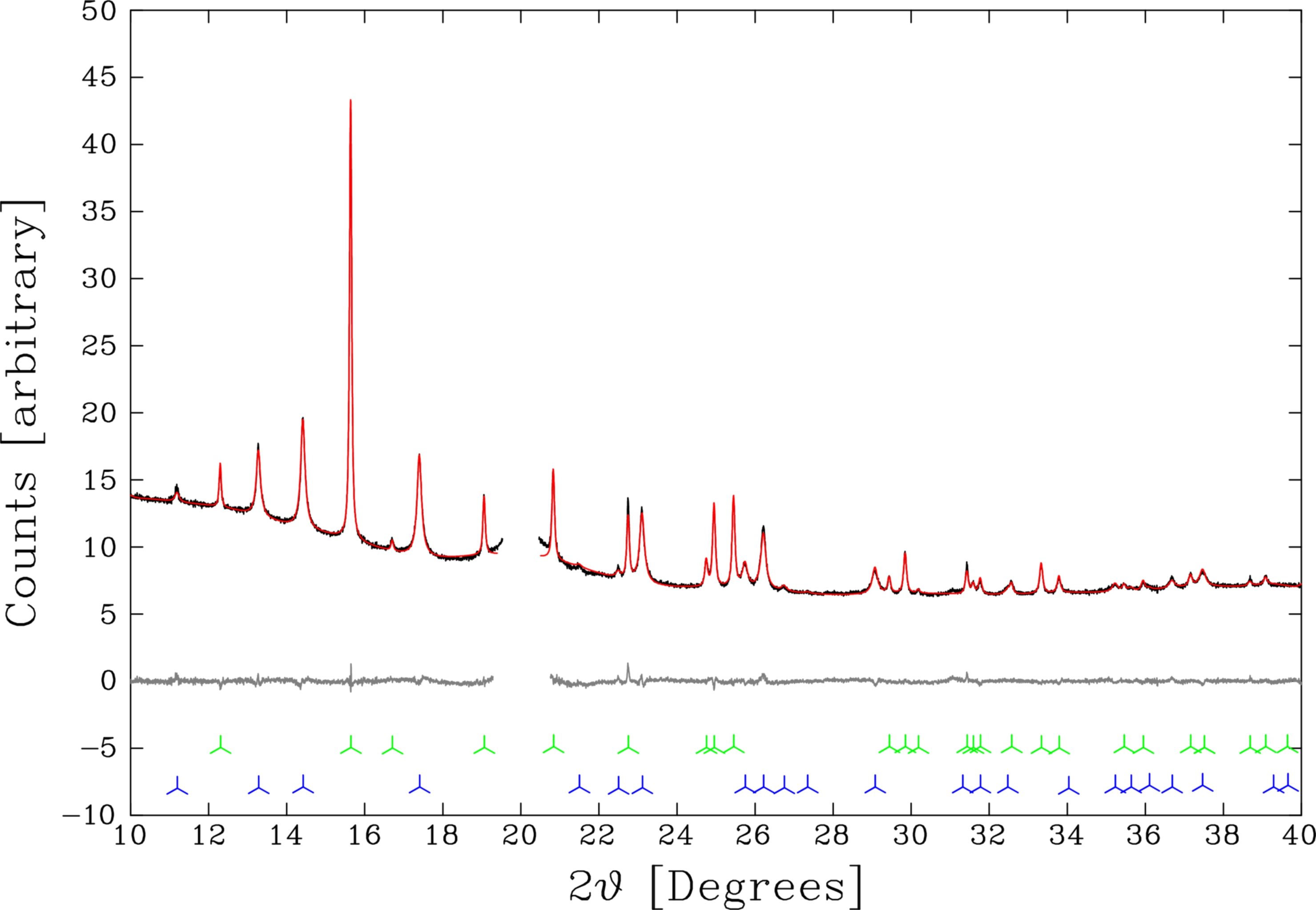} \label{riet122916}}

\subfigure[]{\includegraphics[scale=0.55]{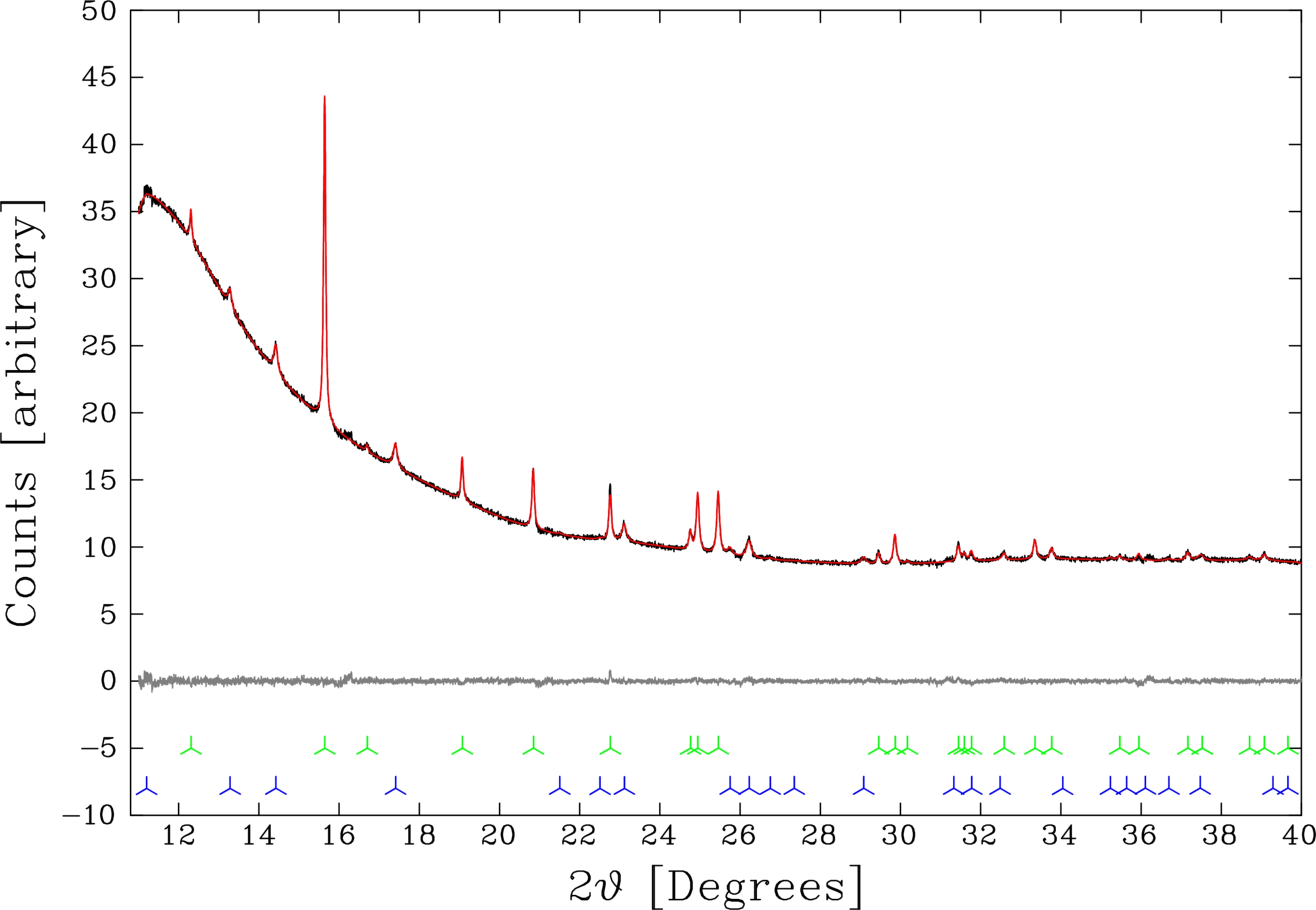} \label{riet123380}}
\caption{Sample rietveld refinements of powder patterns exhibiting vaterite and calcite peaks.  The red curves show the fit, the grey curves show the difference curve and the markers indicate the positions of the individual vaterite (blue) and calcite (green) peaks. a) Initial sample exhibiting vaterite features in sapphire capillary. The sapphire capillary exhibits a strong peak at 19.97\dgr \tth which has been removed from the plot. b) Powder pattern exhibiting vaterite peaks at a CO$_2$ pressure of 30 bar in quartz capillary. \label{rietveld}}

\end{figure*} 
\end{center}

\subsection{Pressure Dependence}

Due to the strong features in the diffraction patterns caused by the use of a sapphire tube it was decided to perform further experiments using quartz-glass capillaries. As before, the sample of amorphous CaSiO$_3$ was initially exposed to a CO$_2$ pressure of 6 bar; however, scans indicated that no major structural changes occurred upon exposure to the gas. The pressure was held at 6 bar for 10 minutes, during this time two very sharp, intense peaks appeared at 16.72\dgr and 33.32\dgr \tth. These correspond to Bragg reflections of calcite, however the intensities are much higher than expected, indicating insufficient powder averaging and therefore likely result from the localised carbonation of a small number of isolated particles, some of which happen to lie in the Bragg condition. As further crystallisation occurs, these particles are taken out of the bragg condition and full crystallisation occurs at higher pressures.

The pressure was then increased to 10 bar and held for 15 minutes, during which the features at 16.72\dgr and 33.32\dgr \tth were observed to diminish with time. Before they disappeared completely, the pressure was increased to 20 bar. Initially, there was no observable effect on the structure; however, after approximately 15 minutes at this pressure the steady growth of multiple crystalline features, due to calcite and vaterite phases, was observed. A further increase in pressure up to 30 bar appeared to accelerate the rate of carbonation, increasing the peak intensities, but did not produce any further changes to the phase structure. \fig{quartzpressevo} shows a selection of SXPD patterns as a function of increasing pressure, showing the evolution of the sample.

\section{Analysis}

\subsection{Thermal Evolution of CaSiO3 \label{sapph}}

Figure~\ref{riet122916} shows the result of a Rietveld structure refinement using the TOPAS-Academic software package \citep{coelho07} and published lattice parameters for vaterite and calcite as starting values. The published lattice parameters used in the refinement were obtained from the Inorganic Crystal Structure Database (ICSD) reference database\footnote{FIZ Karlsruhe http://cds.dl.ac.uk/cds/datasets/crys/icsd/\\llicsd.html} and are shown in Table ~\ref{params}. The vaterite phase was fit with the hexagonal pseudo-cell of \cite{kamhi63}, with unit cell parameters of $a'$ = 4.13~\AA\ and $c'$ = 8.49~\AA\ and space group $P$6$_3$/$mmc$. The reduced hexagonal cell structure was chosen over the larger hexagonal structure of \citeauthor{kamhi63} (\citeyear{kamhi63}; see Table~\ref{cryststruct}) as it provided the best fit to our data. The modelled fit is shown in Figure ~\ref{riet122916} along with the difference pattern between the overall fit and the measured data. The strong peak at 19.97\dgr \tth is due to the sapphire capillary and has therefore been excluded from the fit. The two phase fit produced agreement factors of R$_{wp}$ = 1.78 and R$_{exp}$ = 1.05 with a goodness of fit (GoF) of 1.69; the very low R-factors relate to the high background
present in the data.

The refined lattice parameters and the calculated weight percentage of each phase are shown alongside the published values in Table~\ref{params}. The refined parameters do not differ significantly from the reference values.

\begin{table}[h]
\caption{ Published and refined lattice parameters for vaterite and calcite phases. Published values obtained from the ICSD database. Error values stated are for the final figures. Values given are from samples at room temperature and initial formation pressure; 6 bar for sapphire, 20 bar for quartz. \label{params}}
\begin{tabular}{l l c c c}
\multicolumn{5}{l}{Vaterite -- Space Group: P6$_3$/mmc} \\ \hline
 Capillary & & $a'$ (\AA) & $c'$ (\AA) & Wt \% \\ \hline
 & & & & \\
& Published & 4.13 & 8.49 &  \\ 
Sapphire & Refined &  4.1226 (1) & 8.4653 (3) & 65.5(4) \\ 
Quartz & Refined &  4.1227 (4) & 8.462 (1) & 50.2(4) \\ \hline
\end{tabular} \\
\vspace{20pt}

\begin{tabular}{l l c c c}
\multicolumn{5}{l}{Calcite -- Space Group: R-3c} \\
\hline
 Capillary & & $a$ (\AA) & $c$ (\AA) & Wt \% \\ \hline
 & & & & \\
& Published & 5.05 & 17.32 &  \\
Sapphire & Refined & 4.9869 (1) & 17.0492 (4) & 34.4(4) \\ 
Quartz & Refined & 4.9844 (4) & 17.058 (1) & 49.7(4) \\ \hline
\end{tabular}

\end{table}

\begin{figure}[h]
\begin{center}
\includegraphics[scale=0.40]{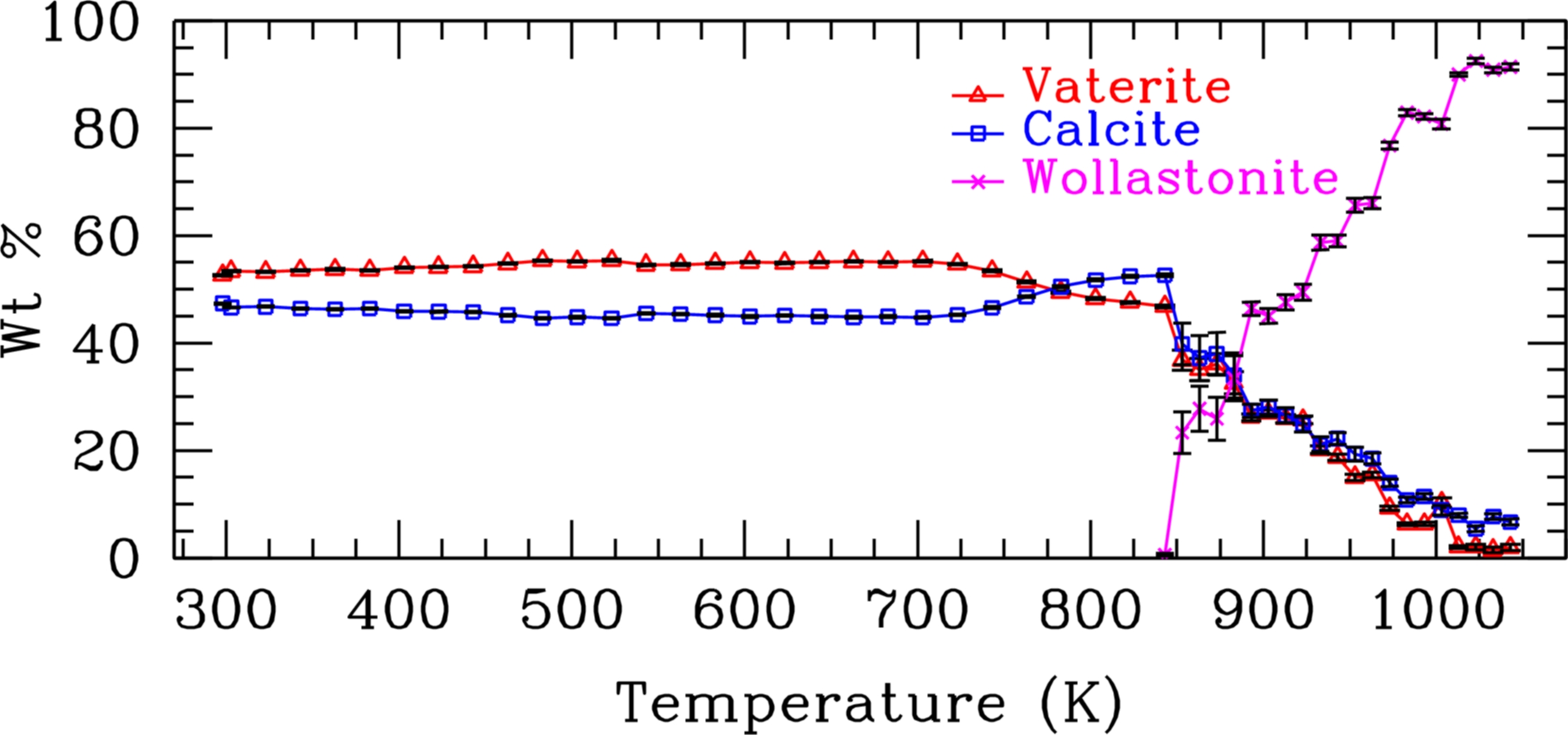}
\end{center}
\caption{Weight percentage against temperature for calcite and vaterite phases of initial CaSiO$_3$ sample in a sapphire capillary, at a pressure of 6 bar. Where error bars are not visible they are smaller than the plotted data points. \label{compsapphire}}
\end{figure} 

Figure ~\ref{compsapphire} is a plot of weight percentage against temperature for the vaterite, calcite and wollastonite phases, with values taken from a batch refinement of selected data sets at approximately 20~K temperature intervals. This provides a more detailed view of what is happening, with regards to composition of the sample at higher temperatures (\textgreater 800~K). From this we can clearly see that, at lower temperatures, the vaterite and calcite phases remain fairly stable, with the relative weight percentage of vaterite increasing by just 3\% over a period of approximately 5 hours. However, once the temperature reaches 723~K the percentage of vaterite begins to fall, being replaced by an increasing amount of calcite. The weight percentage of vaterite continues to fall gradually until calcite supersedes it at a temperature of 770~K. At a temperature of 840~K, wollastonite begins to crystallise and can be included in the fit, this phase then quickly dominates the sample as it continues to crystallise at higher temperatures. \fig{sapphthermalevo} indicates that at a temperature of 973~K, the carbonate peaks have almost disappeared, appearing as weak, broad features, indicating that the carbonate phases are not stable at such temperatures and have broken down, this is also evidenced in \fig{compsapphire} as the amount of vaterite and calcite within the sample at this stage are less than 10 \%. This leads to the sample fully annealing at a temperature of 1103~K to fully crystalline \ca (wollastonite). Alternatively, plotting lattice parameter a against temperature for the vaterite phase indicates a slight thermal expansion between RT and 943~K, this provides a thermal expansion coefficient of $\sim$5$\times$10$^{-5}$K$^{-1}$ (see \fig{vattherexp}).

\subsection{CaSiO$_3$ in Quartz Capillary\label{quartz}}

The crystalline phase formed at a pressure of 20 bar on the second run was identified as also being a combination of calcite and vaterite. Refinement of the data using initial crystallographic parameters obtained from the ICSD database, confirmed this. A typical fit is shown in \fig{riet123380} with the initial and refined parameters and calculated weight percentages listed in Table ~\ref{params}.

\begin{figure}[h]
\begin{center}
\includegraphics[scale=0.4]{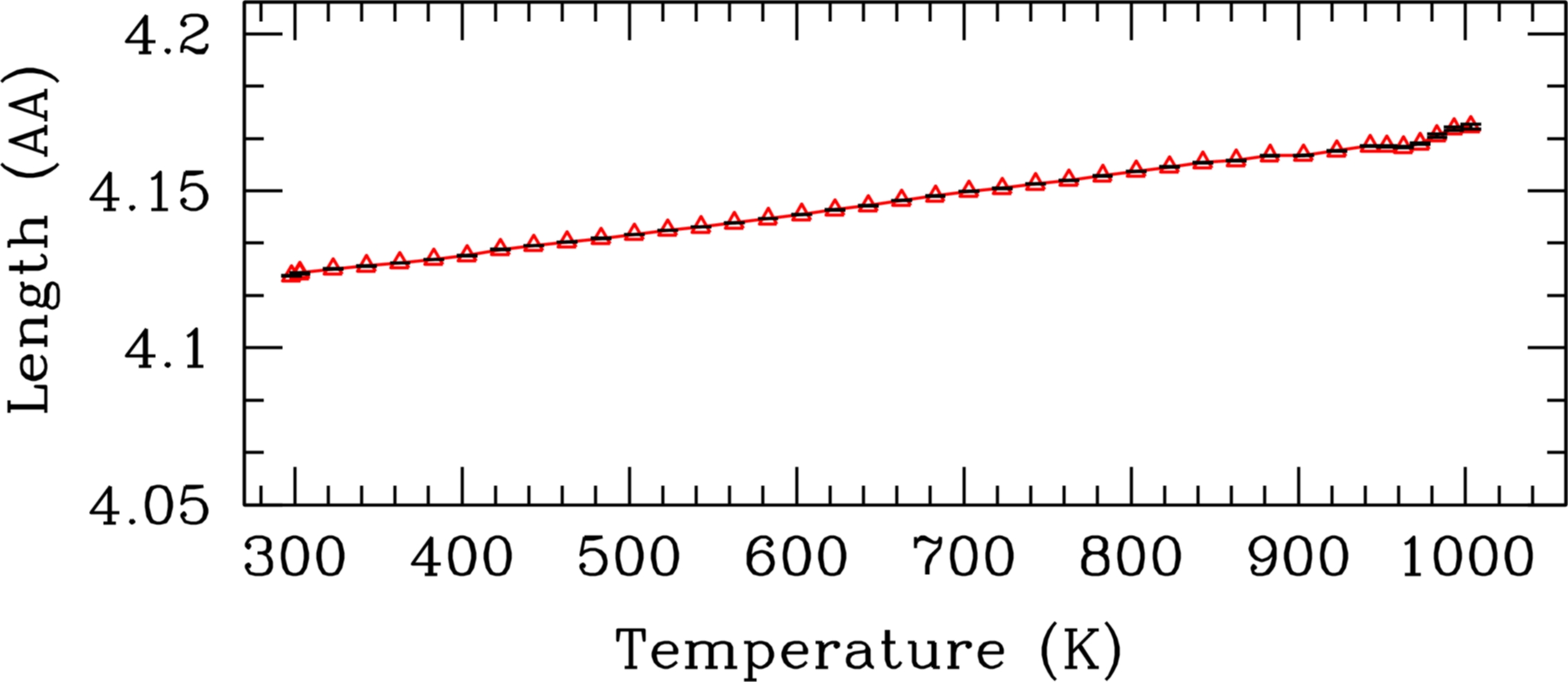}
\end{center}
\caption{The effect of temperature on lattice parameter $a$ for the vaterite phase. Sample in a sapphire capillary at a pressure of 6 bar. \label{vattherexp}}
\end{figure}

The refined values are in good agreement with published values and the refined values for the sapphire capillary sample.
\fig{compquartz} shows the weight percentages of the two phases plotted against pressure and it can be concluded that once the two phases have been formed, they remain relatively stable. An increase in pressure does not appear to have a significant effect on the composition of the two phases.

\begin{figure}[h]
\begin{center}
\includegraphics[scale=0.45]{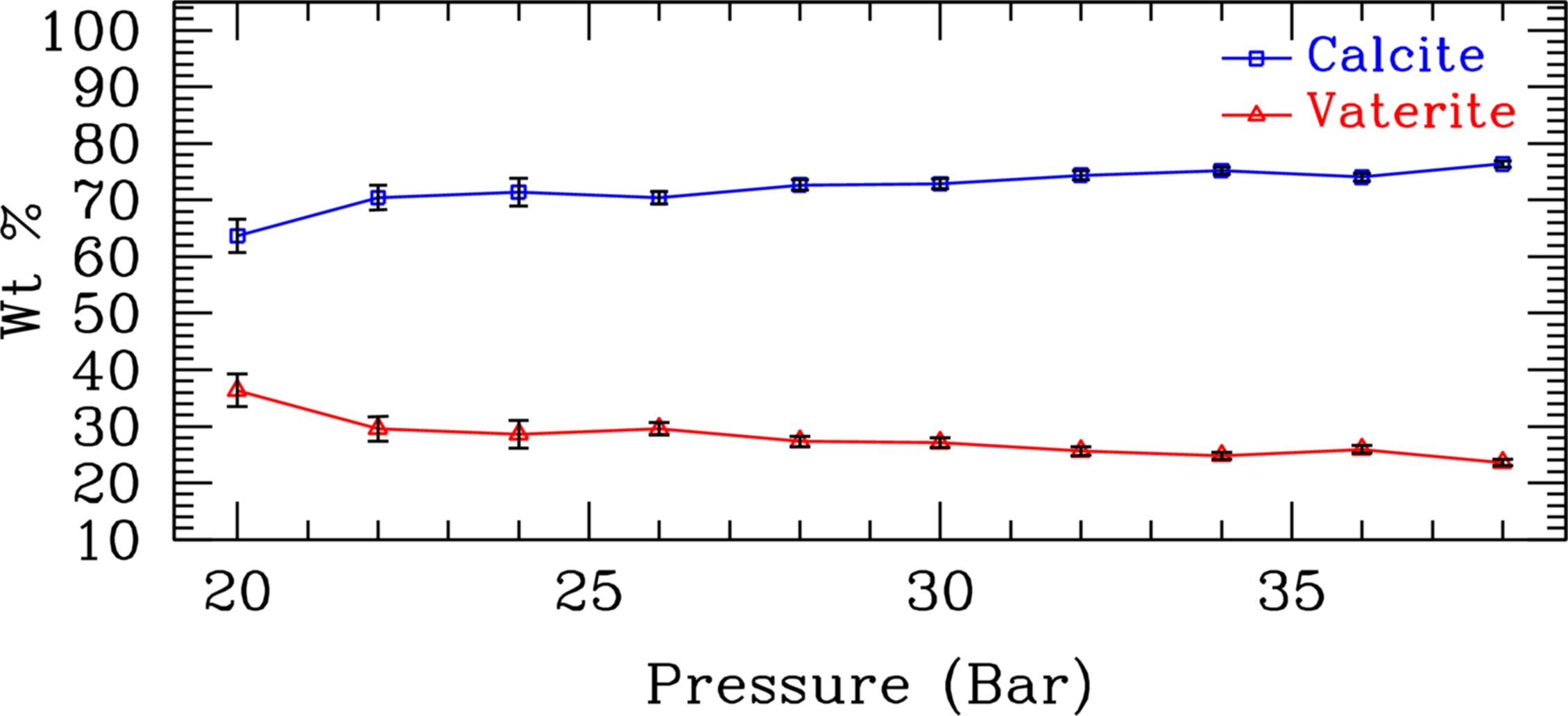}
\end{center}

\caption{Weight percentage against pressure for calcite and vaterite phases at room temperature. Initial sample of CaSiO$_3$ in quartz glass capillary. Where error bars are not evident, they are smaller than the plotted data points. \label{compquartz}}
\end{figure} 

\section{Discussion}

\subsection{Formation of Vaterite}

As discussed in the introduction vaterite is, with respect to aragonite and calcite, the least stable of the anhydrous calcium carbonate phases. Its formation in the present study is therefore interesting, either as a possible component of dust grains, or as a possible metastable precursor phase for the calcite grain component currently attributed to the 92$\mu$m feature \citep{kemper02}. We have shown that the calcium carbonate phase vaterite can be formed through the solid-gas interaction of amorphous calcium-rich silicates and gaseous CO$_2$ at an elevated pressure of 6 bar. 
While the kinetics of the transformation in to calcium carbonate are
dependent on the initial packing density of the powdered CaSiO$_{3}$ sample, 
as discussed in Section 2, the selection of vaterite over the more 
thermodynamically stable calcite is likely to be due to material dependent factors. 
In the following paragraphs we draw on the results of experiments regarding vaterite formation in other 
fields to show that the preferential stabilisation of vaterite in the present case likely arises from 
the presence of SiO$_{2}$ anions.

Due to its industrial and commercial potential, there is a large body of literature that shows vaterite can be stabilized (often temporarily) in the presence of various different ions or organic and inorganic additives under widely differing experimental conditions, most of which are concerned with in vitro formation \citep[e.g.][]{wray57, nassrallah98, nothig99, han06, nebel08, wang09a, fernandez10, gomez10}, or as a post-formation coating \citep{vogel09}. The origin of these stabilisation effects likely lies in the fact that vaterite, aragonite and calcite all have very similar values for their free energy \citep[0.5--3 kJ/mol differences;][]{plummer82, wolf96, wolf00, baitalow98}. Relatively small changes in either the surface area (i.e. small particle size), or in the concentration of impurities, can change their stabilisation properties such that the order of their thermodynamic stabilities becomes inverted \citep{navrotsky03, navrotsky04}. Similar thermodynamic crossovers are observed for several metal oxide systems, including the TiO$_{2}$ polymorphs \citep[rutile, brookite, and anatase;][]{ranade02} and the iron oxide phases \citep{navrotsky08} and can mean that the least stable phases, in the present case vaterite, are often preferentially formed. However, many experimenters (see papers cited above) have reported that, although outcomes are highly reproducible, the actual outcome is also highly dependent on changes in conditions in a non-systematic way \citep[e.g.][]{pai09, ren09}, suggesting that kinetic factors may also be important.
It is also relevant to note Ostwald's Law of Phases, which states that, due to the lower energy barrier of more disordered states, the pathways to final crystallisation of the most stable phase will pass through all less stable phases in order of increasing stability.

For precipitating carbonates a hydrated amorphous calcium carbonate phase (ACC) acts as a precursor, which can then crystallise to a more stable form. Carbonate formation via this route is believed  to follow: ACC $\rightarrow$ vaterite $\rightarrow$ aragonite $\rightarrow$ calcite \citep[e.g.][and references therein]{rodriguez11, bots12}. In this instance, it is believed that due to the local coordination environment of Ca in the nanoframework structure of ACC being similar to those in the crystalline
CaCO$_{3}$ phases \citep{becker03}, the dehydration and condensation of the ACC structure results in the formation of vaterite without major structural rearrangements \citep{gebauer10, goodwin10, rodriguez11}. In the case of carbonate formation by biomineralisation, this often occurs in the presence of silicates and \cite{gal10} have suggested that a mechanism of geometric frustration may operate, whereby the presence of the tetrahedral silicate ion in the flat carbonate lattice hinders the organisation of the carbonate into crystalline polymorphs; while \cite{lakshminarayanan03} previously observed a significant expansion of the calcite lattice with added silicate anion concentration, attributing this to the incorporation of monomeric, oligomeric, and polymeric forms of silicate anion species into the calcite lattice. Such incorporations would increase the calcite lattice energy, making it less stable and therefore less favourable.

Calcium carbonate formation in the presence of various silicates and oxides was also investigated by \cite{lin09}, who identified a correlation between the surface charge provided by the silicate/oxide and the choice of
phase stabilised. Negative surface charge favoured the metastable vaterite and aragonite phases, while positive
charges produced calcite. In crystalline silicates, the possible charge sites are limited by the crystal structure: planar regions which have a negative structural charge and edge regions whose charge arises from the protonation/deprotonation of surface hydroxyls at the side of stacking structures (sheets, chains etc.). \cite{lin09} found amorphous SiO$_{2}$ did not follow the surface charge rule and always stabilised calcite. However they
attributed this to the gel phase that forms when SiO$_{2}$ is exposed to water, suggesting that the hydrated porous nature of the gel is not compatible with epitaxial growth and therefore no templating effect is exerted on carbonate nucleation, allowing calcite to form rather than vaterite or aragonite.

Assuming all the Ca atoms eventually participate in CaCO$_{3}$, for carbonate formation to proceed via the gas-solid reaction in CaSiO$_{3}$  necessitates the formation of SiO$_{2}$, providing a free ionic species, while the absence of liquid water will negate the formation of a gel phase, allowing SiO${_2}$ tetrahedra to either be partially incorporated into the the carbonate structure, as suggested by \cite{gal10} and \cite{lakshminarayanan03}; or, to exert a templating surface charge effect on the forming carbonate as suggested by \cite{lin09}. Since carbonation will occur at the Ca sites within the silicate structure, the initial small size of the carbonate could also contribute to the carbonate free energy to produce a Navrotsky stability inversion and similarly favour the stabilisation of vaterite. The fact that some calcite is also initially formed which, in the second experiment, was observed to be unstable possibly points to contributions such as particle size and remaining silicate ``impurities'' also influencing phase stabilisation priorities. We also acknowledge that kinetic factors may play a role in determining vaterite stabilisation and subsequent growth, particularly as our initial gas injections were performed at room temperature and relatively high gas pressures to overcome sample packing density. However these effects are also likely to favour the formation of disordered structures, i.e. vaterite rather than aragonite or calcite.

The vaterite formed in both cases appears to be of the same structure, both refine well using the hexagonal pseudo-cell values of \cite{kamhi63}. Table ~\ref{params} lists the refined lattice parameters for vaterite at room temperature from both sources. The values are in good agreement with each other, suggesting that, although formed independently, the structure of the vaterite is reproducible.

\subsection{Relevance for Cosmic Dust}

A band near 90$\mu$m was first discovered by \cite{kemper02} in the spectrum of the PNe NGC 6302, followed by detections in the class 0 proto-star IRAS 16293 and two HII regions \citep{ceccarelli02, takashi03}; a subsequent survey of low mass proto-stars showed the feature to be common in such regions \citep{chiavassa05}. The circumstellar feature at $\sim$92$\mu$m was attributed by \cite{ceccarelli02} to calcite, who ruled out the metastable aragonite phase on the basis that its 92$\mu$m feature is too weak and that its formation nominally requires high pressures and temperatures.  Whether vaterite exhibits an observable Far Infrared (FIR) feature, or indeed could be a possible candidate for the carrier of the 92$\mu$m feature, is difficult to assess. There are, to our knowledge, no published FIR spectra for vaterite \citep[e.g review of carbonate spectroscopy by][]{ brusentsova10}, likely made difficult due to a combination of its low thermal stability in respect of producing polyethelene pellets and the difficulty of producing or obtaining pure specimen. However the lattice mode vibrational region is accessible by Raman spectroscopy. The Raman spectroscopic signatures of carbonates are found in three wavenumber regions: 1500--1000, 1000--500, and 500--100 cm$^{-1}$ \citep[e.g.][]{scheetz77, edwards05}. In general, bands at frequencies above 500 cm$^{-1}$ are due to the internal motions of the molecular carbonate ion (internal modes), and those below 500 cm$^{-1}$ are due to motions involving the entire unit cell (lattice modes). In this region the carbonate polymorphs exhibit distinct and characteristic spectral features \citep[e.g.][]{ edwards05, carteret09}. In particular vaterite exhibits a strong broad feature at $\sim$ 100 cm$^{-1}$ and is illustrated in Figure \ref{vat_raman}, which shows Raman spectra for a number of different vaterite samples produced via a biomimetic stabilisation process \citep{thompson11a, thompson11b} that involved the hydrolysis of urea in the presence of amino acid. In these preparations varying quantities of vaterite and aragonite were precipitated along with a minor quantity of calcite. Each polytype is normally distinguishable by its crystal morphology: needles for aragonite, cubes for calcite and spherules for vaterite. For the amino acid, leucine, aragonite exhibited as splintered or branched needles, calcite as stepped/intersecting cuboids and vaterite as flower-like platelets. By focussing the laser spot on the different morphological structures, pure-phase Raman spectra were obtained (Horiba LamRam 800, 532 nm NdYAG laser, 600 line grating, 10$\times$ objective). In the figure, spectra are shown for five samples precipitated under differing conditions. As can be seen in Figure~\ref{vat_raman}, the Raman spectra for these vaterites show a strong feature in the region of 110 cm$^{-1}$($\sim$91$\mu$m).

\begin{table}
\caption{Factor group analysis and selection rules for calcite \citep[after][]{decius77}.}            
\label{table:2}      
\centering                          
\begin{tabular}{c c c c c c c}        
\hline\hline                  
$D_{3d}$ & $A_{1g}$ & $A_{2g}$ & $E_{g}$ & $A_{1u}$ & $A_{2u}$ & $E_{u}$ \\ \hline
Ca $2b$  & 0        & 0        & 0       & 1        & 1        & 2 \\
C $2a$   & 0        & 1        & 1       & 0        & 1        & 1 \\
O $6e$   & 1        & 2        & 3       & 1        & 2        & 3 \\ \hline
$N_{T}$  & 1        & 3        & 4       & 2        & 4        & 6 \\
$T_{A}$  &          &          &         & 0        & 1        & 1 \\        
$T$      & 0        & 1        & 1       & 1        & 1        & 2 \\
$R$      & 0        & 1        & 1       & 0        & 1        & 1 \\
$\Gamma^{vib}$ & 1  & 1        & 2       & 1        & 1        & 2 \\ \hline
Activity & Raman    &          & Raman   &          & IR       & IR \\ 
\hline                                 
\end{tabular}
\tablefoot{ $N_{T} =$ Total number of unit cell modes; $T_{A}=$ acoustic branch modes; $T=$ optic branch translatory modes; $R=$ librational modes; $\Gamma^{vib}=$ internal modes.}
\end{table}

\begin{table}
\caption{Factor group analysis and selection rules for vaterite lattice modes \citep[after][]{anderson91}.}            
\label{table:3}
\centering                          
\begin{tabular}{c c c c c c}        
\hline\hline                  
               & $A_{2u}$ & $E_{1u}$ & $A_{1g}$ & $E_{1g}$ & $E_{2g}$ \\ \hline 
$\Gamma^{vib}$ & 1 & 4 & 2 & 2 & 4 \\
$R$            & 1 & 2 & 1 & 4 & 2 \\
Activity       & IR & IR & Raman & Raman & Raman       \\ 
\hline                                 
\end{tabular}
\tablefoot{$R=$ librational modes; $\Gamma^{vib}(=\Gamma^{vib}_{\mbox{Ca}}+\Gamma^{vib}_{\mbox{CO}_{3}})=$ internal modes.}
\end{table}

   \begin{figure}
	\includegraphics[height=2.5in,width=3.4in]{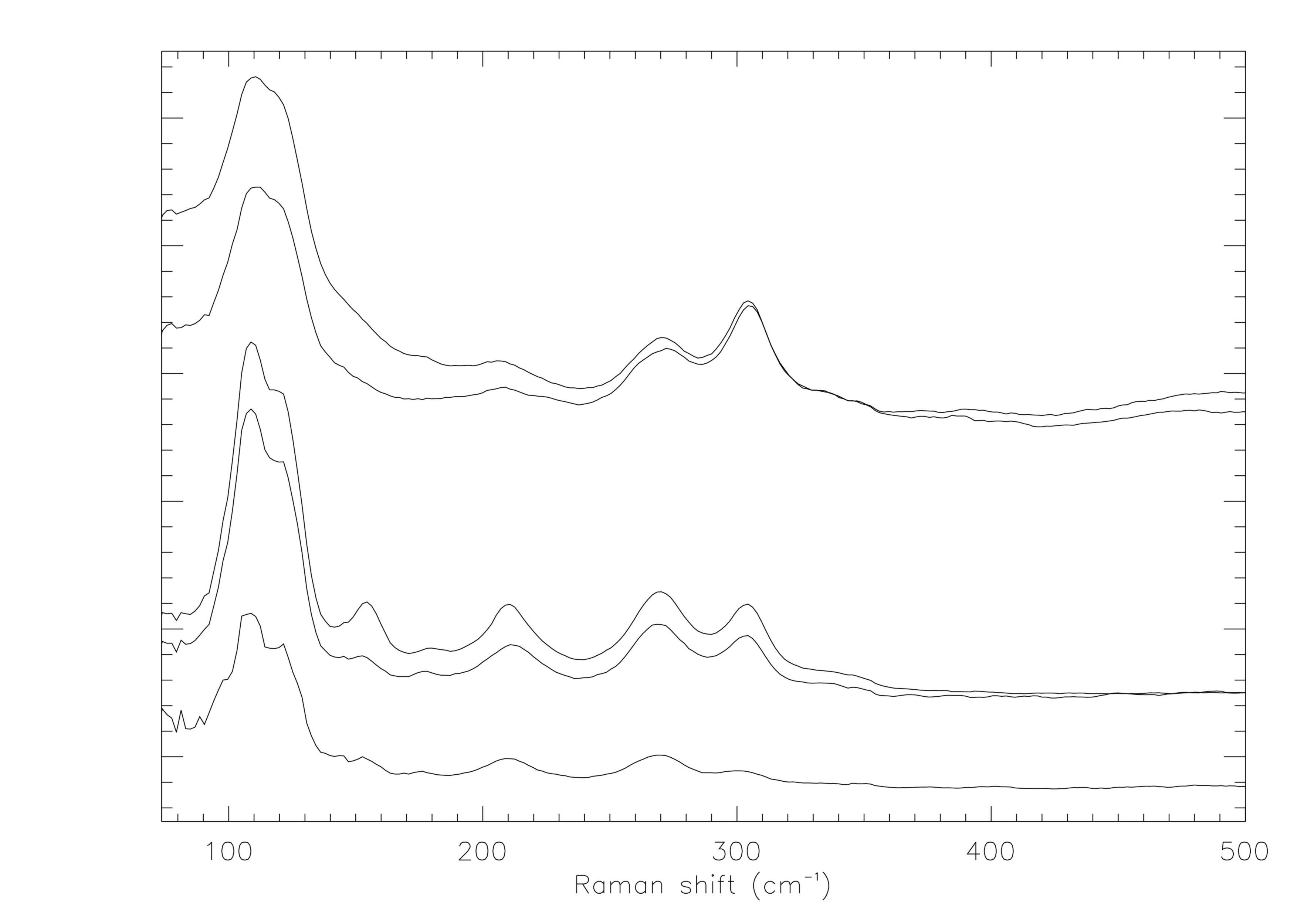}
	\caption[]{Raman spectra for samples of CaCO$_{3}$ vaterite produced by biomimetic synthesis, showing a strong lattice mode feature at $\sim$ 100 cm$^{-1}$. Spectra offset in y-axis direction for clarity and y-axis scale is arbitrary.}
         \label{vat_raman}
   \end{figure}

The lattice-mode assignment for calcite is well established \citep{decius77} with the selection rules shown in Table \ref{table:2}. The coupling which leads to factor-group selection rules is relatively weak such that lattice modes of a given basic origin (e.g. libratory) will be found in similar positions in both IR and Raman spectra \citep{adams80}. In addition, structural disorder, which shows up as broadening of Raman feature widths in the lattice mode region, is a symmetry reducing effect which can in principle lead to a relaxation of the Raman and IR activity selection rules such that certain bands can become both IR and Raman active \citep[e.g.][]{white05, hernandez97, duley05, hopkinson08, abe11}. In the case of vaterite, the uncertainty regarding its unit cell structure and the level of disorder make factor group analysis less certain. However, using the \cite{meyer69} $P6_{3}/mmc$ space group, \cite{anderson91} identified the selection rules listed in Table ~\ref{table:3} which suggests there are eight IR active and 15 Raman active vaterite lattice modes. Based on assignments to measured bands observed at higher frequencies Anderson and Bre\u{c}evi\'{c} predict three IR active bands (corresponding to librations $A_{2u}+2E_{1u}$) should lie below 200 cm$^{-1}$.

Although none of the forgoing is a ``proof-positive'' argument that vaterite possesses a FIR band near 92$\mu$m, it does suggest that measurement of vaterite in the FIR should be addressed as a matter of priority, particularly as the disorder inherent in the vaterite structure might also provide a natural mechanism to account for the observed variations in the carbonate feature width and position.

Although the operational pressures required for the present work arise from the need to provide adequate diffusion of CO$_2$ along the length of the sample capillary, we believe similar interactions involving silicates in circumstellar outflows could occur at much lower pressures. However, even if this is not the case the formation of carbonate via this route may still be directly relevant to proto-stellar and proto-planetary envionments, where carbonates formed at higher pressure on, or in, planetesimal objects could be released as dust grains via disruptive collisional events.

\section{Conclusions}
We have observed the formation of the rare calcium carbonate polymorph vaterite through the solid-gas interaction of amorphous silicates, produced as cosmic dust analogues, and gaseous CO$_2$ at pressures \textgreater 6 bar. In-situ synchrotron X-ray powder diffraction was used to observe the structural evolution of the material during its formation and subsequent heating to a temperature of 1173~K. We determined a range of stability between 6 -- 30 bar and 298~K -- 923~K, a much wider range than previously seen for vaterite. 
This has relevance to the surface compositions of Venus and ancient Mars \citep{vanberk12}, where the CO$_2$ pressure at the surface is sufficiently high for this kind of reaction to occur. Vaterite has been regarded as important due to the fact that it is rarely formed naturally and therefore its presence is usually taken as being indicative of biogenic processes.

Our alternative, inorganic method provides evidence that vaterite found in Martian meteorites and on the surface of Venus and Mars are not necessarily proof of bioactivity but could have formed naturally through a processes similar to those that we describe here.

\section*{Acknowledgements}

The authors would like to thank Prof. Chiu Tang and Mr Jonathan Potter for assistance on the I11 beamline. We thank Dr. J. Nuth for helpful comments on an earlier version of this paper. This work was supported by the Diamond Light Source through beamtime award ee-6515. SJD acknowledges financial support from Diamond Light Source and Keele University. 

 \bibliographystyle{aa}
  \bibliography{vaterite}{}

\end{document}